\documentclass[twocolumn]{aastex631}
\pdfoutput=1 
\usepackage{appendix}
\usepackage{natbib}
\usepackage{graphicx}
\usepackage{enumitem}
\usepackage{hyperref}
\usepackage{mathtools, amsmath,amstext}
\usepackage{cleveref}
\usepackage[T1]{fontenc}
\usepackage[figure,figure*]{hypcap}

\newcommand{\G}{\mathcal{G}}

\newcommand{\appropto}{\mathrel{\vcenter{\offinterlineskip\halign{\hfil$##$\cr\propto\cr\noalign{\kern2pt}\sim\cr\noalign{\kern-2pt}}}}}

\graphicspath{{./}{figures/}}

\submitjournal{ApJ}
\accepted{July 28, 2023}
\shorttitle{Similarity of Near-Resonant Planets}
\shortauthors{Goyal, Dai, \& Wang}

\begin{document}

\title{Enhanced Size Uniformity for Near-resonant Planets}

\author[0000-0001-9652-8384]{Armaan V. Goyal}
\affil{Department of Astronomy, Indiana University, Bloomington, IN 47405}

\author[0000-0002-8958-0683]{Fei Dai} 
\affiliation{Division of Geological and Planetary Sciences,
1200 E California Blvd, Pasadena, CA, 91125, USA}
\affiliation{Department of Astronomy, California Institute of Technology, Pasadena, CA 91125, USA}

\author[0000-0002-7846-6981]{Songhu Wang}
\affil{Department of Astronomy, Indiana University, Bloomington, IN 47405}

\correspondingauthor{Armaan V. Goyal}
\email{armgoyal@iu.edu}



\begin{abstract}
\noindent

Super-Earths within the same close-in, compact planetary system tend to exhibit a striking degree of uniformity in their radius, mass, and orbital spacing, and this \textit{peas-in-a-pod} phenomenon itself serves to provide one of the strongest constrains on planet formation at large. While it has been recently demonstrated from independent samples that such planetary uniformity occurs for both configurations near and distant from mean motion resonance, the question thus remains if the strength of this uniformity itself differs between near-resonant and nonresonant configurations such that the two modes may be astrophysically distinct in their evolution. We thus provide in this work a novel comparative size uniformity analysis for 48 near-resonant and 251 nonresonant multi-planet systems from the California Kepler Survey (CKS) catalog, evaluating uniformity both across systems and between planetary pairs within the same system. We find that while multiplanet configurations exhibit strong peas-in-a-pod size uniformity regardless of their proximity to resonance, near-resonant configurations display enhanced intra-system size uniformity as compared to their analogous nonresonant counterparts at the level of both entire systems and subsystem planetary pairs and chains. These results are broadly consistent with a variety of formation paradigms for multiple-planet systems, such as convergent migration within a turbulent protoplanetary disk or planet-planet interactions incited by postnebular dynamical instabilities. Nevertheless, further investigation is necessary to ascertain whether the nonresonant and near-resonant planetary configurations respectively evolve via a singular process or mechanisms that are dynamically distinct. 
\end{abstract}



\section{Introduction} \label{sec:intro}

The assessment of trends in the architectures of extrasolar planet systems is paramount to the development of a generalized understanding of their evolution, as the emergence of phenotypic patterns over a large sample of systems may provide the initial basis for characterizing possible differences in their formation channels or assembly mechanisms. The most prototypical sample for such an assessment is found in a particularly dominant subset of the mission data from NASA’s \textit{Kepler} Space Telescope \citep{borucki}: the hundreds of systems with multiple close-in ($P \lesssim 100$ days) planets with sizes between that of Earth and Neptune ($1 R_{\oplus} \lesssim R_{p} \lesssim 4 R_{\oplus}$) on low-eccentricity, coplanar orbits ( \citeauthor{xie} \citeyear{xie}; \citeauthor{thompson} \citeyear{thompson}; \citeauthor{millholland2021} \citeyear{millholland2021}). In addition to representing the most common types of planetary systems in our galaxy, these multiplanet super-Earth (SE) or sub-Neptune (SE) configurations are also characterized by a surprising degree of uniformity in the orbital spacing and size of their constituent planets as compared to purely random expectation \citep{weiss_rad}. This “peas-in-a-pod”-style architectural regularity has been the subject of various series of statistical inquiry, and it has been demonstrated that this physical preference for intra-system size uniformity cannot be explained by detection bias alone \citep{weiss_pet}, and may be recovered via frameworks built upon full forward modeling \citep{he} or complexity theory \citep{gilbert}. 

The extension of peas-in-a-pod uniformity to planetary mass, however, necessitates consideration of orbital resonances due to the distinct observational means \citep{lithwick_wu} by which mass measurements may be obtained for systems near or distant from mean motion resonance (MMR): individual transiting planets in a nonresonant system may have their masses constrained by radial velocity (RV) follow-up (\citealt{batalha}; \citealt{weiss2014}), while planet pairs whose orbits lie near MMR may have their masses determined instead via analysis of their transit timing variations (TTVs; \citealt{lithwick_wu}; \citealt{steffen_2012}; \citealt{xie_ttv}; \citealt{hadden}). Nonetheless, intra-system mass uniformity has been demonstrated independently for near-resonant systems exhibiting strong TTVs \citep{millholland}, predominantly nonresonant systems with RV masses \citep{goyal}, a mixed sample of both system types \citep{wang_2017, otegi}, thereby suggesting that the emergence of peas-in-a-pod regularity persists regardless of proximity to MMR. 


  This ubiquity of peas-in-a-pod uniformity across resonant and nonresonant architectures, based on the assessments of independent samples of each against their own random expectation, thus comprises a first-order trend upon which a more detailed investigation of the relationship between MMR and planetary uniformity may be achieved via direct comparison of the \textit{degree} of planetary uniformity between the two types of configurations. The exploration of such a dichotomy is motivated by the fact that many of the most prominent examples of size uniformity within our solar system may be attributed to configurations in or near MMR, such as the Uranus–Neptune (near 2:1) system as well as the Galilean moons Io, Europa, and Ganymede (in 4:2:1 Laplace resonance). Further extrasolar motivation is provided by the resonant chains of Kepler-60, Kepler-80, Kepler-223, K2-138, TRAPPIST-1, TOI-178, and TOI-1136 (\citealt{gozdziewski}; \citealt{macdonald}; \citealt{mills}; \citealt{luger}; \citealt{christiansen}; \citealt{leleu}; \citealt{dai_1136}), all of which seem to harbor more highly uniform planetary masses than typical Kepler systems \citep{goldberg}. As such, it remains to be seen whether near-resonant and nonresonant planetary configurations represent two astrophysically-distinct modes, and if the former does indeed exhibit a greater degree of planetary uniformity than the latter, even within the context of the same system.

We thus present in this work a novel comparison of intra-system size uniformity between 48 near-resonant and 251 nonresonant multiple-planet systems within the California Kepler Survey (CKS) catalog \citep{weiss_samp}. Taking the adjusted Gini index (\citealt{gini}; \citealt{deltas_2003}) as our primary metric for uniformity, we find that near-resonant configurations display enhanced intra-system size uniformity as compared to their analogous nonresonant counterparts, at the level of both entire systems and individual planetary pairs and chains orbiting the same star. We motivate the choice of our sample and statistical methods in Section \ref{sec2}, conduct our primary analysis in Section \ref{sec3}, and provide subsequent statistical validation and astrophysical discussion of our results in Section \ref{sec4}.
    
\section{Sample and Metrics} \label{sec2}
\subsection{Sample Selection and the CKS Catalog}

The relevant sample for the entirety of this work is 299 multiple-planet SE and SN systems within the California Kepler Survey (CKS) catalog \citep{weiss_samp}, which provides a high-purity collection of uniformly derived planet and host parameters for a sizable subset of \textit{Kepler} systems. The CKS observational program yielded substantial improvements on the stellar parameters for 1305 \textit{Kepler} Objects of Interest (KOIs) via high-resolution spectroscopic follow-up \citep{petigura} with the High Resolution Echelle Spectrometer (HIRES; \citealt{vogt}) at the W.M. Keck Observatory, allowing for the provision of 2025 planet candidates with precise radii and host parameters \citep{johnson}. These measurements were further refined via the consideration of parallaxes from \textit{Gaia} Data Release 2  (\citealt{fulton}; \citealt{gaia}). Following the removal of false-positive transit signals, diluted hosts, grazing ($b > 0.9$) transits, low signal-to-noise ratio (S/N; $\text{S/N} < 10$) transit candidates, and potential eclipsing binaries ($R_{p} > 22.4 R_{\oplus}$) from the CKS-\textit{Gaia} catalog, 892 high-fidelity planet candidates remained as part of 349 multiple-planet systems \citep{weiss_samp}. Of these 349 systems, we excluded from our analysis 50 systems hosting at least one planet larger than Neptune ($R_{p} > 4 R_{\oplus}$), thus arriving at the 299 multiple-planet systems (containing 745 total planets) of SE and SN we consider in this work.

As a result of uniform provision in photometric and spectroscopic data for the CKS hosts provided respectively by \textit{Kepler} and HIRES, the 349 CKS multiple-planet systems, and our subset of 299 systems by extension, comprise a homogeneous sample of planet parameters for which a high degree of statistical integrity may be maintained for a relevant population analysis \citep{weiss_samp}. Such homogeneity eliminates the complex biases that may arise from differences in survey construction, instrumentation, heuristics of analysis,  nonstandard treatment of error, and other such systematic discrepancies that may exist in a heterogeneous sample. The elimination of such biases within this sample, as well as the inherent level of measurement precision on planetary radii and orbital parameters, serves to greatly reduce the amount of statistical noise present in our evaluations of resonance and intra-system size uniformity, thus promoting more effective comparison within our intended population analyses.

\subsection{Resonance Parameter}
We distinguish between the near-resonant and nonresonant subsets of our sample using the resonance parameter $\zeta_{n,j}$ from \citet{fab}:

\begin{equation}
\label{eq1}
    \zeta_{n,j} = (n+1)\left( \frac{j}{\mathcal{P}-1} - \text{Round}\left[ \frac{j}{\mathcal{P}-1} \right] \right),
\end{equation}
where $\mathcal{P}$ is the outer-to-inner period ratio of a neighboring planet pair (always larger than unity), $j$ is the order of resonance under consideration, and $(n+1)$ is the order of the adjacent resonances that set the period ratio tolerance for demarcation of the resonant "neighborhood".


Throughout this work, we will consider only proximity to first-order MMR, as these are the most dominant mode of resonance to which \textit{Kepler} systems lie in proximity \citep{fab}. As such, we follow the prescription of \citet{fab} to assign period ratios to neighborhoods of first-order resonances using the $\zeta_{2,1}$ parameter in particular, for which

\begin{equation}
\label{eq2}
    \zeta_{2,1} = 3\left( \frac{1}{\mathcal{P}-1} - \text{Round}\left[ \frac{1}{\mathcal{P}-1} \right] \right).
\end{equation}

While the $n=1$, $j=1$ case of $\zeta_{1,1}$ may also be used to evaluate proximity to first-order resonance, the rationale for using $\zeta_{2,1}$ is that the boundaries of the resonant neighborhoods will be established by the nearest third-order resonances, and are therefore centered more closely around the first-order MMRs of interest, while second-order neighborhoods are also excluded from consideration by construction; $\zeta_{2,1}$ thus provides more effective constraints on proximity to first-order MMR while simultaneously avoiding the erroneous inclusion of systems in nearby second-order resonances, as may occur with $\zeta_{1,1}$ \citep{fab}. More specifically, for a given first-order MMR of index $i$, $\zeta_{2,1}$ will be equal to 0 at the exact MMR ratio $\mathcal{P} = (i+1)/i$ and will extend to values of $-1$ and 1 at the ends of its resonant neighborhood as defined by the two adjacent third-order resonances at $(3i + 4)$:$(3i + 1)$ and $(3i + 2)$:$(3i - 1)$ (i.e. the 3:2 MMR at $\mathcal{P} = 1.50$ will be bounded by the 10:7 MMR, at $\mathcal{P} \approx 1.43$, and the 8:5 MMR at $\mathcal{P} = 1.6$). We may also see here the general benefit of employing the $\zeta_{n,j}$ formalism instead of a simpler method, such as assessing fractional deviation from integer period ratio: while both methods appropriately scale the size of the resonant neighborhood to the specific resonance being considered, the use of $\zeta_{n,j}$ automatically eliminates contamination from higher-order resonances sufficiently close to the desired MMR, a bias for which simpler methods have no effective correction. 

Given that $\zeta_{2,1} = 0$ at the exact period ratios of first-order MMRs and $|\zeta_{2,1}| \leq 1$, close proximity to first-order resonance is best defined by low values of $|\zeta_{2,1}|$ such that, for some arbitrary value $\zeta_{lim}$, near-resonant systems may be effectively characterized by $|\zeta_{2,1}| \leq \zeta_{lim}$. We establish and motivate our choice of $\zeta_{lim}$ in Section \ref{sec3}, and discuss the impacts of altering this boundary in Section \ref{sec4}.

\begin{figure*}
  \includegraphics[width=\textwidth]{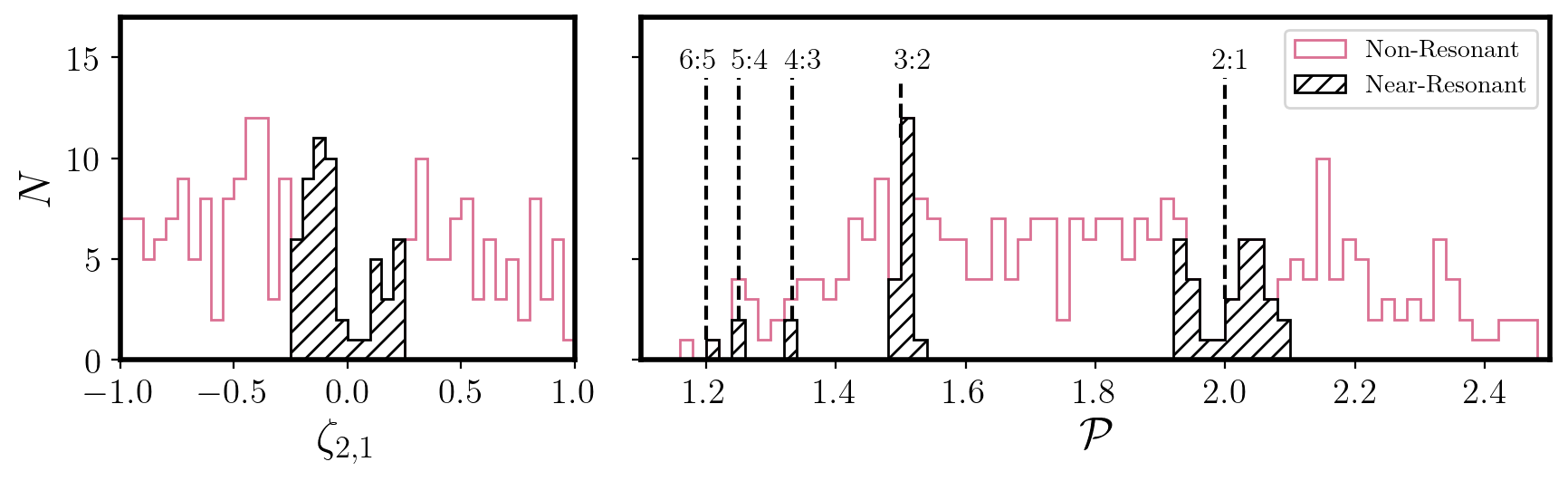}
  \caption{Classification of the 446 neighboring planetary pairs within 299 CKS systems into near-resonant and near-resonant populations. Left: aggregate distribution of the resonance parameter $\zeta_{2,1}$ for the 446 pairs considered in this work partitioned into the 54 near-resonant pairs (hatched black region) selected by $|\zeta_{2,1}| \leq 0.25$ and the remaining 392 nonresonant pairs (pink region). The morphology of our distribution bears a great degree of qualitative similarity to that provided in \citet{fab}, and we observe the same spike at $-0.2 \lesssim \zeta_{2,1} \lesssim -0.1$ that describes an abundance of pairs lying just wide of first-order MMR. Right: direct mapping of our near-resonant and nonresonant classifications onto the period ratio distribution of the 446 CKS pairs. We observe here the effectiveness of our $|\zeta_{2,1}| \leq 0.25$ filter in selecting configurations in close physical proximity to first-order MMR, as well as the natural shrinking of the near-resonant \textit{neighborhoods} as the relevant MMR period ratios approach unity and become increasingly crowded by nearby higher-order resonances (note that the bin width itself is greater than the entire resonant neighborhood for the 4:3, 5:4, and 6:5 first-order modes). This distribution also provides further confirmation that the 2:1 and 3:2 first-order modes are the dominant resonances for \textit{Kepler} systems \citep{fab}. We may also see here more directly the aforementioned asymmetries around each of the MMR integer ratios, for which there is a greater number of pairs that lie just wide of first-order resonances than just narrow of them.}
  \label{fig1}
\end{figure*}

\subsection{Gini Index}

Following the methodology of \citet{goyal}, we adopt the adjusted Gini index (\citealt{gini}; \citealt{deltas_2003}), a statistic commonly used in economics to quantify income or wealth inequality in a given population,  as our primary metric for the assessment of intra-system size uniformity across our sample. For a given data vector $x$ with size $N$, the standard Gini index can be calculated as follows:

\begin{equation}
    G = \frac{1}{2N^{2}\overline{x}}\sum_{i=1}^{N} \sum_{j=1}^{N} |x_{i}-x_{j}|,
\end{equation}
which is equivalent to half of the average difference between all possible pairwise combinations of the data, normalized to the mean $\overline{x}$. The index is unit-normalized such that an ideally uniform dataset would yield $G = 0$, and a maximally diverse dataset would yield $G = 1$. 

However, \citet{deltas_2003} demonstrates the standard Gini index exhibits a substantial bias for which sufficiently small sample sizes ($N \lesssim 10$) will consistently overestimate ($\gtrsim 15\%$ deviation) the true degree of uniformity for a given population, an effect to which planetary systems are highly susceptible since the vast majority of \textit{Kepler} architectures host a number of planets $N_{p} \lesssim 5$. To account for this intrinsic property of the metric, \citet{deltas_2003} also defines an "adjusted" Gini index with an additional corrective prefactor:

\begin{equation}
    \mathcal{G} = \left(\frac{N}{N-1}\right)G.
\end{equation}

We have demonstrated in \citet{goyal} that use of the adjusted Gini index provided by \citet{deltas_2003} effectively neutralizes the aforementioned sample size bias within the same 299 CKS systems utilized in this work, as the mean standard Gini index of 1000 random draws of $N_{p}$ planets exhibits a $\sim 37\%$ variation over $N_{p} \in [2, 5]$, while the mean adjusted Gini is successfully stabilized to within $\sim 2 \%$ across the same range of planet multiplicities. We thus adopt for the entirety of this work the adjusted Gini index for all relevant calculations of planetary size uniformity.

\begin{figure}
  \includegraphics[width=0.45\textwidth]{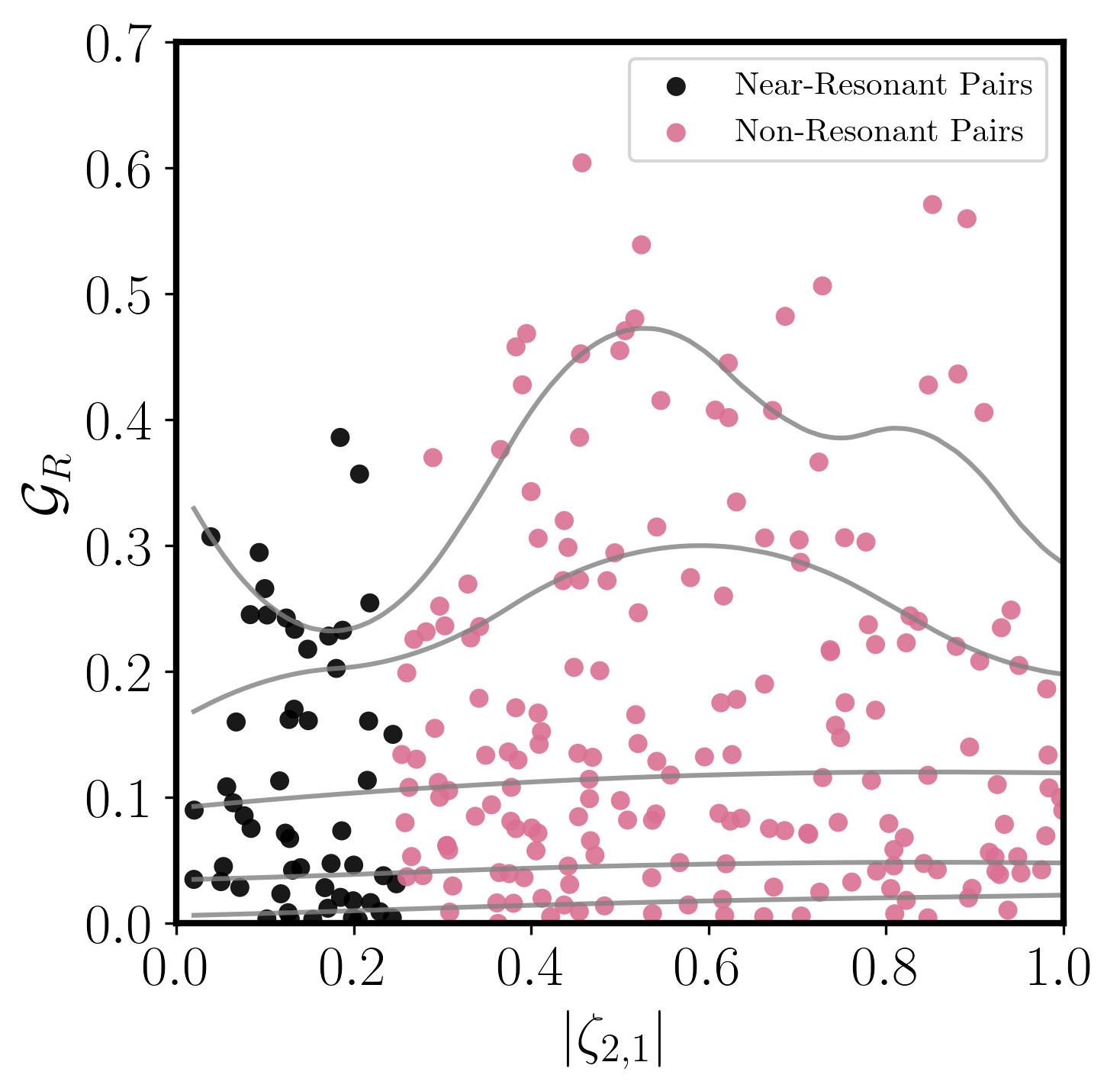}
  \caption{The 54 near-resonant planetary pairs in our sample exhibit an increased propensity for uniform planetary sizes as compared to the 392 nonresonant pairs. 10\% of the nonresonant pairs lie above the maximum Gini value of the near-resonant population ($\mathcal{G}_{R} = 0.38$). The majority ($\sim 57\%$) of near-resonant pairs assume extremely high-uniformity configurations ($\mathcal{G}_{R} \leq 0.1$), while only 43\% of near-resonant pairs occupy the same regime. Gray lines represent smoothed 10th, 25th, 50th, 75th, and 90th percentile regression curves for the global $\zeta_{2,1}-\mathcal{G}_{R}$ sample generated via a constrained B-splines smoothing algorithm (\citealt{cobs1}, \citeyear{cobs2}). We note a pronounced rise in both the 75th and 90th percentile curves with increasing $|\zeta_{2,1}|$, demonstrating a tendency for the most nonuniform pairs to exist farther from MMR, regardless of any binary classification scheme.}
  \label{fig2}
\end{figure}

\section{Analysis and Results} \label{sec3}
\subsection{Pair Classification}
Having established the heuristic basis for the primary metrics within our analysis, we utilize the resonance parameter $\zeta_{2,1}$ to partition our sample of 446 neighboring planetary pairs across 299 multiple-planet systems into near-resonant and nonresonant subgroups. In establishing an appropriate value of $\zeta_{lim}$ for which planet pairs with $|\zeta_{2,1}| \leq \zeta_{lim}$ are considered to be "near" resonance, we find for our sample that a boundary of $\zeta_{lim} = 0.25$ corresponds to a maximum period ratio deviation from first-order MMR of 4.2\% and a median deviation of 1.2\%, thus providing consistency with the near-resonant tolerance level of $0.04\mathcal{P}$ from \citet{fab} while simultaneously imposing a more stringent evaluation of proximity to resonance. 



Applying our $|\zeta_{2,1}| \leq \zeta_{lim} = 0.25$ selection limit to 446 neighboring planet pairs within the 299 systems in our sample, we find that 54 pairs in 48 systems lie sufficiently close to first-order MMR to be considered near-resonant while the remaining 392 pairs comprise the corresponding collection of nonresonant pairs. We illustrate this partitioning of the global $\zeta_{2,1}$ distribution in the left panel of Figure \ref{fig1}, where the pink and black (hatched) portions of each histogram respectively correspond to groupings of near-resonant and nonresonant pairs. We find a notable correspondence in the near-resonant portion of our sample as compared to the analogous distribution from \citet{fab}, where we observe an identical spike from $-0.2 \lesssim \zeta_{2,1} \lesssim -0.1$, which, according to the construction of the resonance parameter given by Equation \ref{eq2}, further confirms the preference for near-resonant Kepler systems to lie slightly wide of their corresponding perfect-integer MMR period ratios (\citealt{delisle}; \citealt{fab}). We map both sets of planet pairs onto period ratio space in the right panel of Figure \ref{fig1}, where the dashed lines indicate the perfect MMR period ratios around which pairs were found in proximity (all first-order resonances up to 6:5). We observe perhaps more readily here the aforementioned preference for asymmetric distributions about the first-order MMR period ratios, as well as the reduction in relative size of the resonant neighborhoods with increasing resonant index (note that the histogram bin width is itself broader than the entire resonant neighborhood for the 4:3, 5:4, and 6:5 MMRs), thus demonstrating the efficacy of the $\zeta_{2,1}$ parameter in assessing close proximity to MMR even for tightly spaced, higher-index resonances.

We calculate the size Gini index for each of the 446 neighboring planet pairs in our sample and assess the behavior of pair size uniformity across the resonance partition illustrated in Figure \ref{fig2}. In performing a nonparametric comparison of the two Gini distributions, we employ the two-sample Anderson-Darling (AD) test due to its greater efficacy in evaluating discrepancies at the distribution tails than the similar two-sample Kolmogorov-Smirnov test. In the context of our sample, the AD test is thereby more sensitive in assessing whether the near- and nonresonant populations differ in their respective propensities for extremely high- or low-uniformity pairs. Applying the AD test to our two distributions, we obtain a probability of $p=0.009$ for emergence from a common parent distribution, indicating a statistically significant level of discrepancy. The near-resonant pairs are entirely contained within $\mathcal{G}_{R} \leq 0.38$, a value above which 10\% of the nonresonant pairs reside. The near-resonant pairs also exhibit specifically strong clustering near the highest-uniformity values of $\mathcal{G}_{R} \leq 0.1$, with 57\% of near-resonant pairs residing within this regime compared to 43\% of nonresonant pairs. Additionally, we calculate the outer-to-inner planetary size ratios for either pair classification, recovering respective distributions of $R_{p, i+1}/R_{p,i} = 1.08 \pm 0.33$ and $1.14 \pm 0.56$ for the near-resonant and nonresonant pairs. We see that, while both median size ratios are slightly larger than unity, illustrative of first-order consistency of either population with observed trends of intra-system size uniformity and size ordering \citep{weiss_rad}, the dispersion of the nonresonant population is $\sim 70\%$ greater than that of its near-resonant analog, suggesting that the latter may harbor a greater occurrence of especially highly uniform planetary pairs.

To perform a more robust and objective assessment of a possible relationship between pairwise size uniformity and distance from MMR, we briefly disregard the imposition of a discrete $\zeta_{lim}$ partition and consider instead a means of fully bivariate evaluation of the $\zeta_{2,1}-\mathcal{G}_{R}$ space across the total pair sample. To remove potential biases that may result from assumptions of Gaussianity in the pairwise Gini distribution or linearity between the two variables under consideration, we conduct a nonparametric analysis of our sample using a constrained B-spline (COBS) smoothing algorithm (\citealt{cobs1}, \citeyear{cobs2}), from which we construct smoothed 10th, 25th, 50th, 75th, and 90th percentile regression curves (Fig. \ref{fig2}, gray lines) for the total $\zeta_{2,1}-\mathcal{G}_{R}$ distribution. While the COBS contours are less suggestive of strong variation in the high-uniformity regime, with the bottom half of Gini distribution generally contained within $\mathcal{G}_{R} \lesssim 0.1$ across the full range of $|\zeta_{2,1}|$, they maintain strong support for discrepancies in the high-uniformity region of the distribution, with the 75th percentile curve for $|\zeta_{2,1}| \gtrsim 0.25$ lying near or above the 90th percentile curve for $|\zeta_{2,1}| \lesssim 0.25$ for a large section of the relevant domain ($0.3 \lesssim |\zeta_{2,1}| \lesssim 0.8$), while the median value of the 90th percentile curve for $|\zeta_{2,1}| > 0.25$ $(\mathcal{G}_{R} = 0.35)$ is nearly identical to the maximum Gini value achieved for $|\zeta_{2,1}| < 0.25$ $(\mathcal{G}_{R} = 0.38)$. Most prominently, we observe that, even in the absence of a defined near-resonant partition within the sample, there still exists a strong qualitative augmentation in the Gini values of the 75th and 90th percentile curves with increasing $|\zeta_{2,1}|$, indicating that, regardless of any prescribed classification scheme, the most nonuniform planetary pairs tend to favor configurations far from MMR.

\subsection{System Classification}
\begin{figure}
  \includegraphics[width=0.45\textwidth]{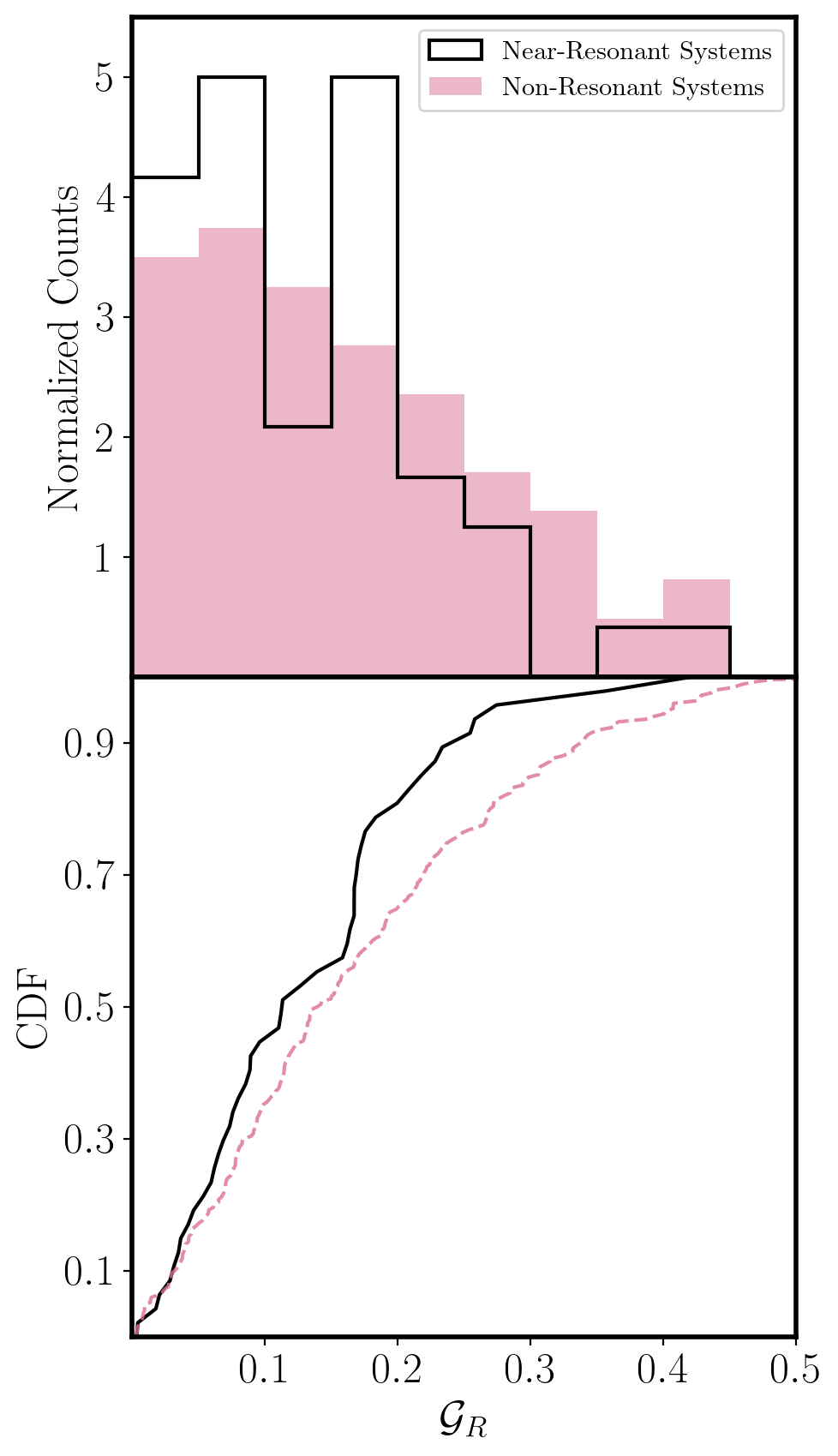}
  \caption{Near-resonant systems exhibit a greater preference for high-uniformity architectures and disfavoring of high degrees of diversity in planetary sizes as compared to nonresonant systems. Top: unit-normalized histograms corresponding to $\mathcal{G}_{R}$ distributions for the near-resonant (black) and nonresonant (pink) populations. We observe both a relative overrepresentation of near-resonant systems in the high-uniformity ($\mathcal{G}_{R} \leq 0.2$) regime as well as the lack of a high-diversity ($\mathcal{G}_{R} > 0.3$) tail that is present in the nonresonant group. Bottom: corresponding $\mathcal{G}_{R}$ cumulative distributions functions (CDFs) for the two populations. The near-resonant systems are 38\% and 81\% contained by respective low-Gini regimes of $\mathcal{G}_{R} \leq 0.1$ and $\mathcal{G}_{R} \leq 0.2$, for which only 28\% and 65\% of nonresonant systems reside. The 75th and 90th percentile Gini values for the near-resonant population ($\mathcal{G}_{R} = 0.17$ and $\mathcal{G}_{R} = 0.24$) also lie well below the respective values for the nonresonant population ($\mathcal{G}_{R} = 0.24$ and $\mathcal{G}_{R} = 0.34$).}
  \label{fig3}
\end{figure}

We hereafter classify an entire planetary system as near-resonant if it contains at least one of the 54 planetary pairs with $|\zeta_{2,1}| \leq 0.25$, and analogously deem a whole system as nonresonant if \textit{all} of its constituent planet pairs have $|\zeta_{2,1}| > 0.25$. We thus sort our 299 CKS systems into 48 near-resonant and 251 nonresonant systems, and plot the corresponding size Gini index distributions (unit-normalized histogram and cumulative distribution function, hereafter CDF) for either population in Figure \ref{fig3}. As performed for the distributions of the individual planetary pairs, we apply the two-sample AD test to the Gini distributions for our near- and nonresonant systems, obtaining a probability of $p=0.079$ for incidence from a common parent distribution. While this value itself falls slightly above the conventional threshold for statistical significance, it is still representative of a level of discrepancy that manifests as readily identifiable qualitative differences between the distributions in question. In similar fashion to the aforementioned dichotomy between near- and nonresonant planetary pairs, we observe immediately that the distribution of near-resonant systems is far more pronounced at the low-Gini end, with 38\% and 81\% of  near-resonant systems contained respectively within $\mathcal{G}_{R} = 0.1$ and $\mathcal{G}_{R} = 0.2$, compared to 28\% and 65\% of nonresonant systems. We note also for the near-resonant distribution the suppression of a high-Gini tail that appears to be present in the nonresonant sample, evidenced by the 75th ($\mathcal{G}_{R} = 0.17$) and 90th ($\mathcal{G}_{R} = 0.24$) percentile Gini indices of the near-resonant systems lying far below the analogous values ($\mathcal{G}_{R} = 0.24$ and $\mathcal{G}_{R} = 0.34$) for the nonresonant systems. We thus recover a tendency for near-resonant planetary systems to provide greater occupation of the low-Gini, high-uniformity regime while exhibiting a markedly lower propensity for highly heterogeneous planet sizes. 

We note that the metrics and statistical results presented thus are intended only as intuitive and heuristically simple suggestive evidence for the heightened propensity of near-resonant planetary pairs and systems to be more uniform in size than their nonresonant counterparts. A more detailed and statistically rigorous investigation of the same trends will serve as the subject of the remainder of this section.

\begin{figure}
  \includegraphics[width=0.45\textwidth]{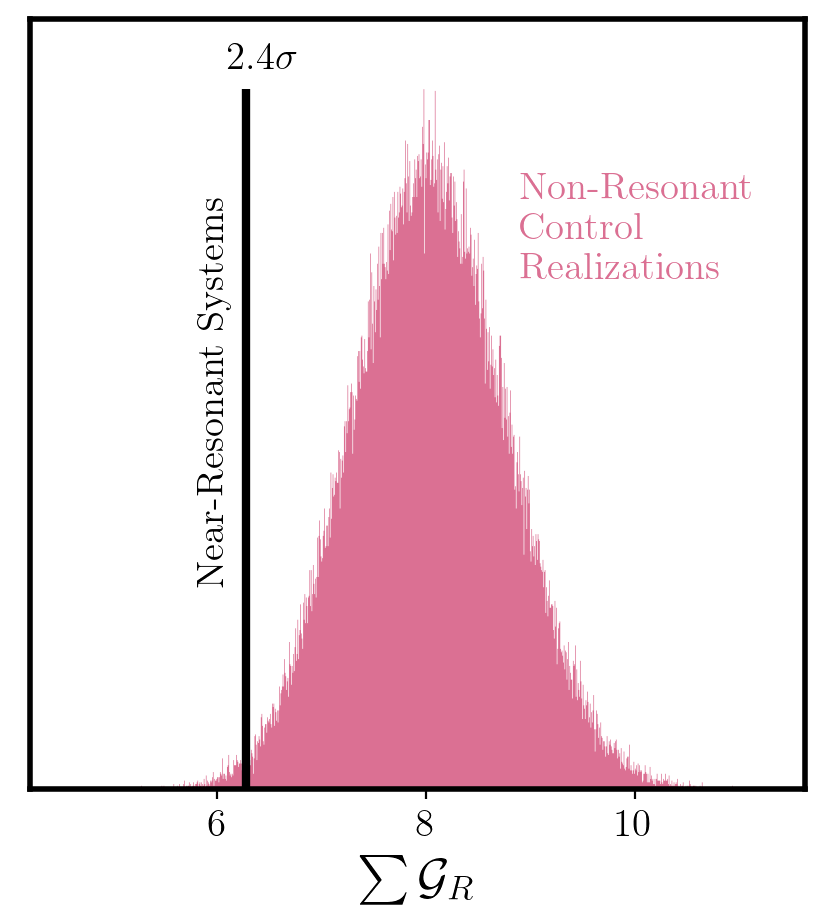}
  \caption{The 48 near-resonant systems in our sample exhibit enhanced intra-system size uniformity at the level of $\sim 2.4 \sigma$ confidence when compared to a distribution of equally sized, randomly drawn mock samples from the nonresonant population. The pink histogram illustrates the control distribution of total size Gini $\sum \mathcal{G}_{R}$ values for $10^{5}$ randomly drawn (without replacement) samples of 48 systems from the 251 in the nonresonant population. The black line denotes the value of $\sum \mathcal{G}_{R}$ calculated for the 48 near-resonant systems, which falls well below the bulk of the nonresonant control distribution and thus denotes a stronger overall expression of intra-system size uniformity within the near-resonant population.}
  \label{fig4}
\end{figure}

\subsection{Size Uniformity across Systems} \label{across}

We shall first verify that the populations of near-resonant and nonresonant systems individually exhibit intra-system size uniformity as compared to their respective random expectation, for which we subject either group to the null hypothesis procedure featured in \cite{goyal}: For a given sample of systems, we compute the adjusted size Gini $\mathcal{G_{R}}$ of each system and compare the sum of these values across our real sample to a distribution of analogous summed Gini values for $10^{4}$ mock samples wherein all individual planetary radii were randomly shuffled (without replacement) across the sample itself. For our 48 near-resonant and 251 nonresonant systems, we find that intra-system planetary size uniformity is present against a null hypothesis of purely random assortment with respective significance values of $6.4 \sigma$ and $4.8 \sigma$, thereby indicating that system-level peas-in-a-pod size uniformity is indeed present within either population.

To ascertain the existence of a possible discrepancy in the strength of this size uniformity between the two populations,  we employ a null hypothesis test similar in construction to the analyses presented by \citet{goyal} and \citet{millholland}. The null hypothesis in question is the assumption that size uniformity is wholly independent of proximity to first-order resonance, and may consequently be assessed via direct comparison of the aggregate uniformity across the near-resonant population to that of random, equally sized subsamples of the nonresonant population. As such, we measure the total size Gini index ($\sum \mathcal{G}_{R}$) for our 48 near-resonant systems against a control distribution of $\sum \mathcal{G}_{R}$ values generated for $10^{5}$ subsamples of 48 systems drawn randomly (without replacement) from the 251 systems in the nonresonant population. 

Illustrating our results in Figure \ref{fig4}, we see that the total size Gini index of the near-resonant population ($\sum \mathcal{G}_{R} = 6.28$, black line) falls well below the vast majority of the values which comprise the nonresonant control distribution ($\sum \mathcal{G}_{R} = 8.03 \pm 0.74$, pink histogram). Consequently, the null hypothesis is rejected at the level of $\sim 2.4 \sigma$ confidence, thus consistent with enhanced intra-system size uniformity for near-resonant systems that holds a $<1\%$ probability of incidence from the nonresonant population.

Given that all near-resonant systems must, by definition, contain at least one planetary pair with period ratio $\mathcal{P} \lesssim 2$, while nonresonant systems have no such restriction, we shall attempt to verify that the dichotomy characterized by our result is not significantly confounded by physical differences between closely-spaced and more distant pairs, perhaps owing to varied degrees of similarity in their formation environment. To this end, we repeat our experiment for only the 110 systems wherein all constituent planetary pairs are at least as closely spaced as the widest near-resonant pair in our sample ($\mathcal{P} < 2.1$). We find that the 35 near-resonant systems therein still display enhanced size uniformity with $2.2\sigma$ significance ($< 1.5\%$ probability of chance occurrence) as compared to the 75 nonresonant systems of similar compactness, thereby affirming that our result is not driven by differences in pair spacing between the two sub-populations.

Given this level of statistical significance that may be attributed to the  enhanced size uniformity of near-resonant systems within the CKS sample, we briefly consider here the possible means by which the observed strength of this trend itself may be bolstered for a larger overall sample. Operating under the assumptions that the intrinsic fraction of near-resonant multiple-planet systems ($f_{res}  \approx 0.17$), as well as the true size Gini distributions for both near- and nonresonant systems, may be modeled adequately by the analogous quantities within the CKS sample thus considered, we attempt to determine the characteristic sample size $N_{3\sigma}$ at which the experimental procedure described in Figure \ref{fig4} would reject its null hypothesis with $3\sigma$ significance. For various trial sample sizes $N_{sys}$, we respectively draw, at random, $N_{sys}f_{res}$ and $N_{sys}(1-f_{res})$ values from the near- and nonresonant CDFs in Figure \ref{fig3} to serve as mock samples for either group, for which we then apply our null hypothesis test to obtain a corresponding significance value. We then perform 1000 bootstrap iterations of this process at each value of $N_{sys}$ and assign to $N_{3\sigma}$ the sample size for which the median significance level is equal to $3\sigma$, obtaining a nominal sample size of  $N_{3\sigma} \approx 440$. While such a $\sim 50\%$ increase in sample size may be obtained, in principle, by the inclusion of other candidate systems to the sample considered here, we shall remark once again that the primary motivation for limiting our analysis to the CKS dataset alone is the especially high degree of statistical purity for the data therein, emergent from an extensive spectroscopic follow-up campaign, refinement with supplemental astrometry, and homogeneity in observational systematics and parameter estimation techniques (\citealt{johnson}; \citealt{petigura}; \citealt{fulton}). As such, the CKS dataset remains at present the most suitable sample for the analyses considered thus, though the provision of a larger sample of multiplanet systems, subject to similarly detailed and homogeneous follow-up observations, may indeed be achieved with the Transiting Exoplanet Survey Satellite, or with the European Space Agency's PLAnetary Transits and Oscillations of stars mission.


\begin{figure*}
  \includegraphics[width=\textwidth]{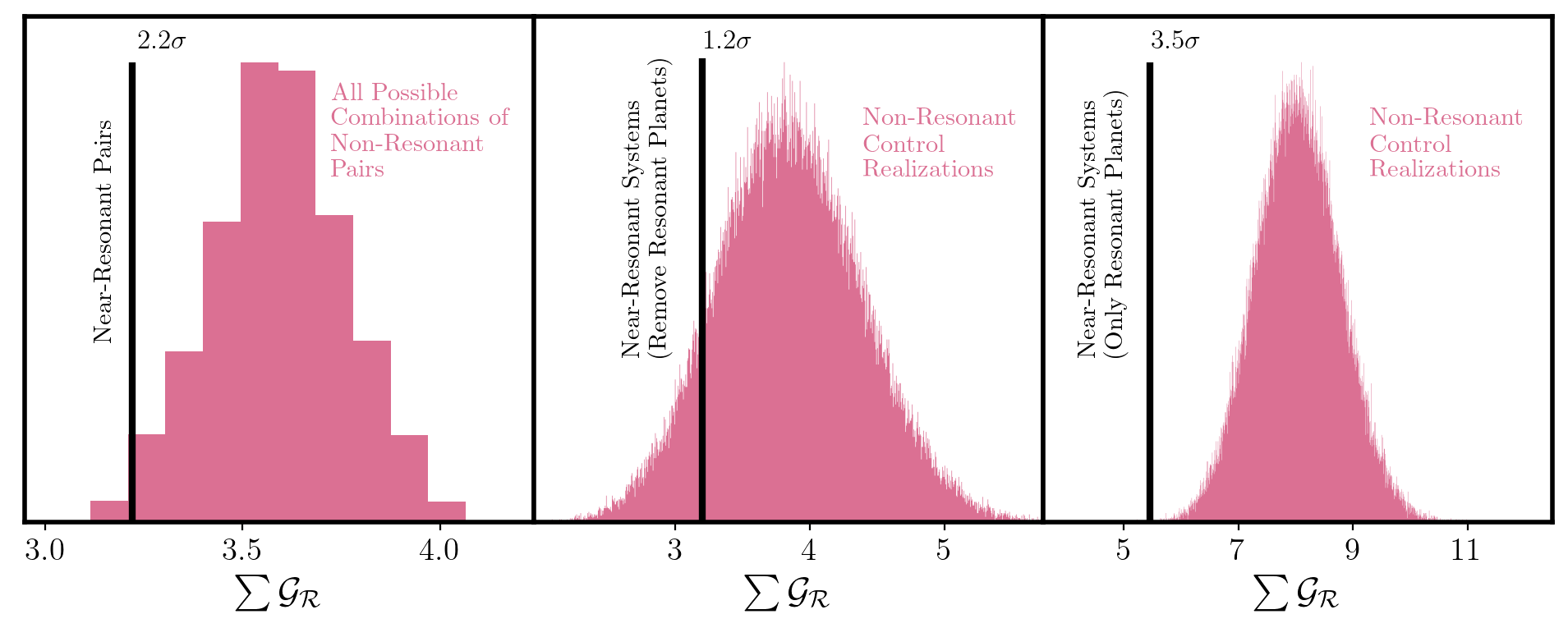}
  \caption{The enhanced intra-system size uniformity exhibited by near-resonant systems is driven primarily by the size similarity of near-resonant pairs and chains. Left: near-resonant pairs exhibit enhanced size uniformity when compared to nonresonant pairs in the same system. Across 27 systems with both near-resonant and nonresonant pairs, the total size Gini of nonresonant pairs (black line) falls $\sim 2.2 \sigma$ below the distribution of total Gini values for equally sized samples of nonresonant pairs drawn randomly each system (pink histogram). Note that the relative coarseness of the histogram is a result of the probability space of random shufflings being fully encapsulated in 2304 possible combinations, far less than the $\gtrsim 10^{5}$ combinations available for our other statistical tests. Center: the nonresonant components of near-resonant systems are statistically indistinguishable from systems that are completely nonresonant. The total Gini index for 23 systems rendered nonresonant via removal of their innermost or outermost planets (black line) lies within $\sim 1.2 \sigma$ of the control distribution generated from $10^{5}$ random draws of 23 nonresonant systems (pink histogram). Right: near-resonant pairs or chains within near-resonant systems exhibit even greater enhancement in their size uniformity than the same systems taken as a whole. The total Gini index of the 48 near-resonant systems in our sample, once removed of all nonresonant pairs (black line), lies $\sim 3.5 \sigma$ below the distribution generated from $10^{5}$ random draws of 48 nonresonant systems (pink historgram).}
  \label{fig5}
\end{figure*}

\subsection{Size Uniformity within Systems} \label{within}
Having established the statistical prevalence of enhanced size uniformity for systems containing at least one near-resonant planetary pair compared to those devoid of such pairs, we now consider potential differences in the degrees of size uniformity for near-resonant and nonresonant substructures within these systems themselves. The question remains of whether the trend elicited by the system-level comparison in Figure \ref{fig4} is the result of enhanced size uniformity for all planets in a near-resonant systems, including its constituent nonresonant pairs, or enhanced size uniformity that is exclusive to near-resonant pairs and chains alone. In order to ascertain which of these modes is dominant, we perform here three modified versions of the null hypothesis testing scheme described in Section \ref{across}.

We first probe the possibility of inherently discrepant degrees of size uniformity between near-resonant and nonresonant planetary pairs within the same system (Figure \ref{fig5}, left panel). Considering from our sample the 27 near-resonant systems with an equal or greater number of nonresonant pairs as near-resonant  pairs, we sum the size Gini indices of the individual resonant pairs within each system, then summing across all 27 systems to obtain a total Gini index for the near-resonant pairs (black line). We then randomly select from each system nonresonant pairs equal in number to the near-resonant pairs of the same system, repeating the summation process to obtain nonresonant control values. This random selection process allows for 2304 possible combinations of nonresonant pairs between the 27 systems considered, the corresponding total Gini values of which serve to comprise our nonresonant control distribution (pink histogram). We see that the total size Gini of the  near-resonant pairs ($\sum \mathcal{G}_{R} = 3.22$) lies below the nonresonant control distribution ($\sum \mathcal{G}_{R} = 3.60 \pm 0.17$) at the $\sim 2.2 \sigma$ level ($<1.4\%$ probability of chance occurrence), providing evidence for the enhanced size uniformity of near-resonant planetary pairs compared to nonresonant pairs within the same system.

We then compare the respective degrees of intra-system uniformity for the nonresonant components of near-resonant systems and systems that are entirely nonresonant (Figure \ref{fig5}, central panel). We determine the nonresonant portions of our near-resonant systems via the removal of individual planets belonging to resonant pairs, and in order to preserve the greatest integrity of extant architectures and prevent the creation of nonphysical gaps in orbital structure, we exclusively consider the systems that lose their resonant signature from the removal of only their innermost or outermost planet(s). This removal process results in 23 systems rendered nonresonant from this truncation, which are then compared to the 251 naturally nonresonant systems in our total sample via the same null hypothesis process described in Section \ref{across}. We see that the total size Gini of the 23 truncated systems ($\sum \mathcal{G}_{R} = 3.20$) displays consistency with the nonresonant control distribution ($\sum \mathcal{G}_{R} = 3.84 \pm 0.54$) within $\sim 1.2 \sigma$, suggesting that the nonresonant components of near-resonant systems do not exhibit any statistically significant enhancement in their size uniformity.

Finally, we perform an experiment complementary to the previous test by assessing intra-system size uniformity for only the near-resonant components of near-resonant systems (Figure \ref{fig5}, right panel). For all 48 near-resonant systems, we remove all nonresonant neighboring pairs and compare the collection of remaining near-resonant pairs or chains in each system to the 251 nonresonant systems via the same null hypothesis test from Section \ref{across}. We find that the total Gini index for these resonant-only configurations ($\sum \mathcal{G}_{R} = 5.46$) lies beneath the nonresonant control distribution ($\sum \mathcal{G}_{R} = 8.03 \pm 0.74$) with $\sim 3.5 \sigma$ confidence ($<0.03\%$ probability of chance occurrence), implying that the near-resonant pairs or chains within a given system exhibit a greater enhancement in their size uniformity than the same system considered in its entirety.

Similar to the heuristic argument presented at the beginning of Section \ref{across}, we subject our 23 systems rendered nonresonant via truncation (\ref{fig5}, center) and our 48 near-resonant systems with nonresonant planets removed (\ref{fig5}, right) to the null hypothesis procedure featured in \cite{goyal} to ensure that either sample maintains intra-system size uniformity as compared to random expectation. We find that the two samples respectively demonstrate system-level peas-in-a-pod size uniformity with $3.8\sigma$ and $6.0 \sigma$ significance, thereby indicating that intra-system size uniformity is indeed maintained regardless of the presence of near-resonant pairs, while the enhancement of this uniformity in a given sample is modulated by such pairs.

We note that recent work by \citet{millholland_split} has demonstrated the existence of intra-system architectural substructure amongst planets on either side of the radius valley, where isolated treatment of either SE or SN within a given system yields size uniformity twice as strong as that displayed by the system considered globally. Accordingly, we wish to ascertain if the differences in pairwise size uniformity thus presented are themselves subject to confounding effects from these \textit{split peas-in-a-pod} architectures, the most prominent evidence for which would present as a higher intrinsic rate of homogeneity in classification (SE-SE or SN-SN) for near-resonant pairs as compared to nonresonant pairs. As performed in \citet{millholland_split}, we classify each planet in our sample as a SE or SN based on its residence below or above the radius valley boundary parameterized by \citet{van_eylen_radius_gap} as $\log_{10} (R_{p}/R_{\oplus}) = m \log_{10} (P\text{days}) + a$, with $m = -0.09^{+0.02}_{-0.04}$ and $a = 0.37^{+0.04}_{-0.02}$. We perform 1000 bootstrap iterations of this classification, where the values of $m$ and $a$ are resampled uniformly within their $1\sigma$ errors. Across this process, we find that $68.5^{+5.6}_{-2.1}\%$ of the near-resonant pairs and $62.5^{+2.1}_{-1.5}\%$ of the nonresonant pairs are of homogeneous planetary type. Since these rates are consistent within the $2\sigma$ level, even when uncertainties are limited to the empirical scatter of the classification scheme itself, we affirm that the enhancement in near-resonant pairwise size uniformity presented in this work is not a byproduct of an augmented rate of homogeneity in planetary type, and is most likely driven primarily by proximity to first-order MMR. 

Taken together, the results of the tests presented in Figures \ref{fig4} and \ref{fig5} may be best summarized as the following: although multiplanet systems demonstrate intra-system size uniformity regardless of their proximity to resonance, near-resonant systems display an enhancement in their size uniformity that is primarily driven by enhanced similarity of planetary pairs and chains close to first-order MMR. We expound upon the astrophysical underpinnings of these findings in Section \ref{theory}, specifically in terms of their implications for the dynamical evolution of multiple-planet systems.

\section{Discussion} \label{sec4}
\subsection{Statistical Validation} 
Our primary results thus strongly support the notion that near-resonant and nonresonant configurations are distinct in their provision of planetary size uniformity, but given that these results are inexorably linked to certain assumptions and prescriptions inherent to the construction of our analysis, it remains to be seen whether the emergence of this dichotomy is truly indicative of physical differences between the two populations, and is not itself a statistical artifact or product of bias. As such, we shall perform here a series of additional tests to demonstrate that our methodology is statistically and heuristically robust, and to further validate the notion that the emergence of enhanced size uniformity is primarily dependent upon physical proximity to first-order MMR. 

\begin{figure*}
  \includegraphics[width=\textwidth]{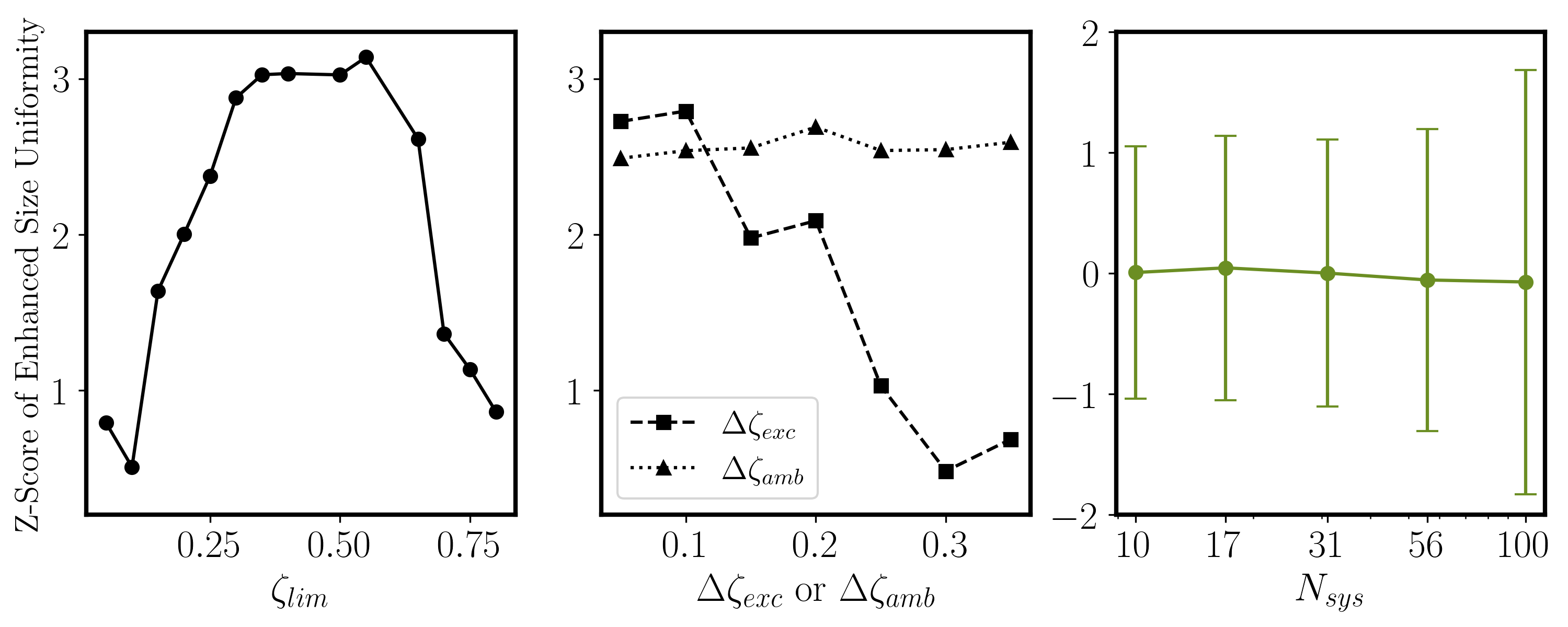}
  \caption{Additional statistical evidence supporting the notion that enhancement in size uniformity for near-resonant systems is driven primarily by inherent physical differences between the near-resonant and nonresonant populations. Left: we observe from variation in the near-resonant cutoff parameter $\zeta_{lim}$ the behavior expected from a physical link between intra-system size uniformity and proximity to resonance: low significance values for incomplete near-resonance populations ($\zeta_{lim} < 0.2$), followed by an increase and subsequent plateau as more resonant systems are added to this population ($0.2 \leq \zeta_{lim} < 0.55$), and terminated by a drop in significance as the added \textit{near-resonant} systems are sufficiently deviant from MMR to no longer exhibit enhanced size uniformity ($\zeta_{lim} \geq 0.55$). Middle: The significance of enhanced size uniformity drops steadily as the near-resonant population is offset progressively from MMR ($\Delta \zeta_{exc}$ increases, dashed line), while being maintained at or beyond the $2.4 \sigma$ level as the boundary between the near- and nonresonant populations is given increasing nonzero width $\Delta \zeta_{amb}$ (dotted line). Both of these trends are suggestive of a physical link between the prevalence of enhanced size uniformity and proximity to MMR. Right: significance distributions for 1000 iterations of the null hypothesis test where considerations of resonance are entirely ignored, and the \textit{near-resonant} population is instead $N_{sys}$ systems drawn randomly (without replacement) from the overall CKS population. We observe that values for all $N_{sys}$ are consistent with null significance and exhibit no apparent trend with sample size, thus suggesting our primary $\sim 2.4 \sigma$ result cannot be obtained by random population assignment and that our null hypothesis procedure is itself largely devoid of underlying sample size biases.}
  \label{fig6}
\end{figure*}

\subsubsection{Prescription of Near-resonant Classification Boundary}
The first such test consists of repeating the null hypothesis procedure illustrated in Figure \ref{fig4} for different values of the $\zeta_{lim}$ parameter, the prescription of which functions as the primary selection criterion for the near-resonant population (Section \ref{sec2}). If there indeed exists an inherent physical link between size uniformity and proximity to first-order resonance, one may expect that the statistical significance of the discrepancy between near-uniform and nonuniform systems to behave in a rather predictable manner with regard to changes in the boundary between the two populations. Low significance values ought to occur for the low-$\zeta_{lim}$ regime since the near-resonant population for such tight limits may still exclude the various systems close enough in proximity to MMR to exhibit enhanced size uniformity. The significance values may then be expected to rise with the broadening of $\zeta_{lim}$ as more of these enhanced systems are included in the near-resonant population, before experiencing a plateau then decreasing at sufficiently high values of $\zeta_{lim}$ for which systems included in the near-resonant population are now distant enough from MMR such that they no longer exhibit enhanced size uniformity. We thus perform the null hypothesis test from Section \ref{across} for various trial values of $\zeta_{lim}$, plotting the significance results of each iteration in Figure \ref{fig6} (left panel). From these tests, we observe behavior that is entirely consistent with our aforementioned qualitative expectations, with the near-resonant and nonresonant populations being indistinct at low values of $\zeta_{lim}$, the former exhibiting a stable but significant level of enhanced size uniformity at values beyond our prescribed cut of $\zeta_{lim} = 0.25$, and returning to insignificant levels of discrepancy for sufficiently large values ($\zeta_{lim} \gtrsim 0.55$). We shall note here that, while the $0.3 \lesssim \zeta_{lim} \lesssim 0.55$ regime is consistent with a plateau at $\sim 3\sigma$, thus representative of an even more significant dichotomy between the near-resonant and nonresonant populations than our $\sim 2.4\sigma$ result with $\zeta_{lim} = 0.25$, we nonetheless maintain the latter limit due to its aforementioned consistency with the $0.04 \mathcal{P}$ limit for near-resonance provided by \citet{fab}, a figure with which values of $\zeta_{lim} = 0.3$ (maximum deviation of $0.052 \mathcal{P}$) or larger are not compatible.

\subsubsection{Influence of Systems Closest to MMR}
A corollary of this test, which is itself predicated on adjusting the scheme of classification for near-resonant systems, is to also repeat the procedure from Section \ref{across} with \textit{near-resonant} populations that are progressively offset from $|\zeta_{2,1}| = 0$, and are thus exclusive of the systems that exhibit the greatest degree of proximity to first-order resonance. We introduce an offset parameter $\Delta \zeta_{exc}$, such that the near-resonant population is now defined by $\Delta \zeta_{exc} \leq |\zeta_{2,1}| < 0.25 + \Delta \zeta_{exc}$, and repeat our primary null hypothesis procedure for various trial values of $\Delta \zeta_{exc}$ (Figure \ref{fig5}, middle panel). We observe a pronounced decrease in the significance of enhanced size uniformity with increasing offset of the near-resonant population from MMR, thereby further motivating the intrinsic relationship between proximity to first-order MMR and the emergence of highly uniform planetary architectures.

\subsubsection{Influence of Ambiguously Classified Systems}
Having now investigated the behavior of the enhanced size uniformity as subject to alterations in the classification of the near-resonant population, we shall now perform a similar assessment with regard to changes to the classification of nonresonant systems. It is arguable that the imposition of a single, discrete boundary value $\zeta_{lim}$ to separate near-resonant and nonresonant systems may lead to ambiguous classification for those systems near the boundary itself, as those systems with planetary pairs just beyond the value of $\zeta_{lim}$ may perhaps hold configurations that are more physically akin to members of the near-resonant population. In order to account for such ambiguous cases and to ensure that the systems within the nonresonant population are all sufficiently distant from MMR to be architecturally distinct from the near-resonant systems, we consider an additional parameter $\Delta \zeta_{amb}$ that describes the width of an \textit{ambiguity zone} within which systems are disregarded from classification altogether, such that the near-resonant population maintains its inclusion criterion of $|\zeta_{2,1}| \leq \zeta_{lim}$ but the nonresonant population is now defined instead with $|\zeta_{2,1}| > \zeta_{lim} + \Delta \zeta_{amb}$. We repeat our null hypothesis testing for various ambiguity widths $\Delta \zeta_{amb}$ (Figure \ref{fig6}, middle panel), recovering significance values of $2.6 \sigma \pm 0.2 \sigma$ for each value considered. As such, we find that the additional limitation of the nonresonant population to only those systems most distant from MMR served to consistently maintain or only mildly increase the statistical dichotomy present between the two groups, thus furthering the notion that the provision of enhanced size uniformity is a distinct physical property exclusive to those systems closest to MMR.

\subsubsection{Random System Classification and Sample Size Effects}
Finally, in order to provide further confirmation that the observed significance of enhanced size uniformity for near-resonant systems cannot be achieved by a purely random classification scheme, as well as to verify that the significance values reported for the suite of tests presented in this work are not strongly influenced by sample size biases, we now conduct our primary null hypothesis scheme for complementary populations whose assignment is purely random and entirely independent of resonance. As such, we shall assign a given sample size $N_{sys}$ for our \textit{near-resonant} population for which we will draw $N_{sys}$ random systems (without replacement) from the 299 CKS systems, thus allowing the \textit{nonresonant} control population to simply comprise the $299-N_{sys}$ systems that remain. We shall repeat this random sampling procedure 1000 times for a given sample size so as to generate various mock populations to which our null hypothesis test can be applied, thus achieving a distribution of significance values expected for randomly determined \textit{near-resonant} populations of size $N_{sys}$. We consider first the case of $N_{sys} = 48$ to maintain consistency with the real near-resonant population determined by the $|\zeta_{2,1}| \leq 0.25$ criterion employed in our primary analysis. We find that 1000 mock samples of $N_{sys} = 48$ drawn randomly from the 299 CKS systems achieve a significance distribution of $0.02 \sigma \pm 1.20 \sigma$, suggesting that our $\sim 2.4 \sigma$ primary result thus holds only a $\lesssim 1\%$ probability of occurrence from a classification scheme that is agnostic to resonance. We also repeat this process for trial sample sizes of $N_{sys} \in [10, 17, 31, 56, 100]$, chosen to be logarithmically equidistant nodes between $10$ and $10^{2}$, and compare the resulting significance distributions of each in Figure \ref{fig6} (green points and bottom axis). We see immediately that each of these distributions are centered very close to null significance, thus confirming the base expectation that such \textit{near-resonant} and \textit{nonresonant} populations should be statistically indistinguishable, regardless of sample size, if both are constructed in a purely random manner from the aggregate CKS population. This ubiquitous consistency with null significance, along with the lack of any emergent correlation with sample size, also demonstrates that any significance values obtained from the null hypothesis procedure in this work are largely free of sample size biases and that the strength of our statistical results are therefore primarily driven by inherent physical differences in size uniformity.

\subsection{Astrophysical Implications} \label{theory}
Having thus validated the statistical prevalence of enhanced intra-system size uniformity for near-resonant pairs and systems as compared to their nonresonant counterparts, we shall now attempt to provide insight towards the dynamical and evolutionary mechanisms out of which the observed architectural phenomena may emerge for either configuration.

\subsubsection{MMR as a Consequence of Planet-Disk Interactions}
It has been demonstrated thoroughly that the formation of close-in, resonant chains may be a preferred  outcome of dynamical evolution within the protoplanetary disk \citep{terquem}, as viscous accretion gives rise to convergent planetary migration toward the inner edge of the disk itself, from which further planet-planet interactions and gas-mediated energetic dampening lead to the highly efficient capture of planetary pairs or chains into first-order MMR (\citealt{morrison}; \citealt{broz}). This notion of a nearly-ubiquitous resonant history for multiple-planet systems implies that the crux of their architectural evolution may lie with the transition of such configurations from being in resonance to near-resonant or from either to entirely nonresonant, and whether such pathways are themselves distinct or exist synergistically.

The signature of any generalized dynamical history for near-resonant configurations is most perhaps most distinctly preserved in the aforementioned overabundance of such pairs at period ratios $\sim 1\%$ wide of first-order MMR (\citealt{fab}, Figure \ref{fig1} of this work). It has been reasoned that these modal near-resonant configurations may be achieved as a result of planet-disk interactions during the migration phase \citep{choksi}, where such broadened spacing may itself be an equilibrium state modulated by disk properties \citep{terquem2} or the result of stochastic turbulence induced by fluctuations in disk density \citep{batygin_adams}. Alternative models posit that this broadening may occur following the disk epoch. It has been demonstrated analytically that the presence of weak postnebular eccentricity damping begets the loss of orbital energy as heat, which naturally allows for a repulsion between planetary pairs that is itself strongest near resonant commensurabilities and weakens away from MMR, thus promoting the existence of the observed quasi-stable near-resonant configurations (\citealt{lithwick_wu}; \citealt{batygin2013}). The requisite weak eccentricity damping for such a process may be provided from host-mediated tidal circularization \citep{delisle}, obliquity tides from neighboring planets \citep{millholland3}, or interactions with a pervading field of planetismals \citep{chatterjee}. 

\subsubsection{Turbulent Disk Migration: Quiescent Disruption of MMR}
Recent work by \citet{goldberg2} has demonstrated that, of these various paradigms, the turbulent disk model is particularly amenable to reproducing the properties of observed planetary configurations, as it provides energy dissipation sufficient to reproduce the aforementioned degree of broadened spacing while simultaneously introducing stochastic forces that excite the mixed resonance angles into circulation, thereby successfully dislodging planets from MMR on both accounts. In consideration of planetary pairs within such a disk, we note that, since the provision of nearly equal masses constitutes a stable and energetically optimal configuration for a two-planet system (\citealt{adams2}; \citealt{adams1}), it reasons that invariance to such stochastic perturbations may itself scale with planetary uniformity as well, such that highly uniform pairs may have a greater tendency to retain their orbital spacing while less uniform pairs may drift farther from MMR \citep{adams1}, where subsequent secular interactions between pairs significantly dislodged from MMR may further sculpt their orbital evolution. Accordingly, the enhanced uniformity of near-resonant configurations presented in this work is a natural consequence of such a turbulent disk model, as is the departure of planetary pairs from their nascent resonant state. 

Despite the efficacy of this model in these regards, it has been remarked that the efficiency of disk turbulence in dissolving mean motion commensurabilities depends sensitively on both the planet-to-star mass ratio of a given system as well as the local surface density of the disk itself \citep{batygin_adams}, such that the resulting stochastic forces are unlikely to serve as the universal means of repulsion and excitation out of MMR. Consequently, even in regimes where this dissolution is favorable, perturbations induced by disk turbulence may not be sufficiently disruptive to uniquely generate the nonresonant configurations that dominate the observed sample, indicating that such modal architectures may be the result of complementary or alternative mechanisms. Furthermore, while the degree of stochasticity afforded by turbulence may indeed disrupt idealized capture into MMR, it may be too greatly amenable to the narrowing of pair spacing (\citealt{goldberg2}; \citealt{choksi2}) in a manner that disfavors the well-characterized excess of pairs lying slightly wide of commensurability (Figure \ref{fig1}). It has been demonstrated that such asymmetries in the period ratio distribution may be preserved under a scheme with comparatively smoother migration torques, while mixed resonant angles for a given pair may still be excited into circulation via secular forcing from an external perturbing planet \citep{choksi2}.

\subsubsection{Postnebular Dynamical Instabilities: Violent Disruption of MMR}
It has thus been posited recently that nonresonant configurations may be achieved readily via a ``breaking the chains” pathway in which postnebular dynamical instabilities, generated by gravitational dynamics, rapidly rotating host stars (\citealt{spalding1}; \citealt{spalding2}), or system mass loss \citep{matsumoto}, give rise to planet-planet interactions that dislodge individual planet pairs from MMR and serve to reshape the planetary mass distribution via collisions and mergers \citep{izidoro}. \citet{goldberg} demonstrate through dynamical simulations that the emergence of such instabilities within resonant chains may successfully reproduce the observed characteristics of multiple planet systems: using the present-day resonant chains Kepler-60, Kepler-80, Kepler-223, K2-138, TRAPPIST-1, and TOI-178 (\citealt{gozdziewski}; \citealt{macdonald}; \citealt{mills}; \citealt{luger}; \citealt{christiansen}; \citealt{leleu}) as models for the initial postnebular architectures of their synthetic systems and triggering instability via a 10\% mass reduction for all planets, \citet{goldberg} illustrate that the subsequent dynamical evolution of such systems yields a smooth period ratio distribution and degrees of mass uniformity, spacing uniformity, and coplanarity that are all in statistical agreement with the analogous qualities of the observed nonresonant population. Most notably, this paradigm provides an efficient mechanism to resculpt nonresonant planetary pairs out of their primordial degree of uniformity, allowing for the widespread achievement of the statistical phenomena treated in this work. 

However, as encountered in the case of the turbulent disk model, this formalism alone may not suffice as a global framework for the evolution of multiple-planet systems. The postnebular instabilities characteristic of this model are expected to occur efficiently and near-ubiquitously ( \citealt{ghosh}; \citealt{lammers}) such that only a few percent of systems would be expected to retain their initial resonant signatures \citep{goldberg2}, which lies in direct tension with the observed occurrence of near-resonant planetary configurations. While the achievement of near-resonant structures in such a formalism could thus occur primarily by chance following the epoch of instability \citep{goldberg2}, such near-resonant pairs would lose memory of their primoridal uniformity and lack any means of subsequent coordination in planetary size, disallowing the prospect of enhanced size uniformity altogether.

\subsubsection{Planetary Density as an Evolutionary Diagnostic: Revisiting the RV vs. TTV Mass Discrepancy}
The two models presented here thus comprise an apparent schism in their establishment of the observed architectures of near-resonant and nonresonant planetary configurations, where disk turbulence may be too quiescent a process to generate the prevalence of the latter while postnebular collisions and mergers may be too violent to preserve the stability of the former. Nonetheless, evidence for the possible coexistence of these mechanisms may itself be treated empirically via density measurements of nonresonant planets in systems with and without additional near-resonant pairs: if planets in entirely nonresonant systems indeed experienced more disruptive postnebular dynamical evolution, it reasons that nonresonant planets in such systems would have lost their primordial massive atmospheres while nonresonant planets in near-resonant systems may have retained their large gaseous envelopes (\citealt{inamdar}; \citealt{biersteker}). As such, the provision of lower mean density values for the latter type of planets as compared to the former would validate the notion of the aforementioned distinct formation paradigms for near-resonant and nonresonant systems. The existence of such a dichotomy may also inform a potential astrophysical basis for the observed discrepancies in the density distributions of TTV and RV planets, where the former comprise systematically higher mean density values than the latter (\citealt{weiss2014}; \citealt{hadden}). Although this discrepancy may owe its nature in part to reduced transit probability and photoevaporation efficiency for small planets at long orbital periods \citep{gaidos}, inherent heuristic differences in TTV and RV detection biases \citep{steffen}, or systematic underestimation of TTV amplitudes from fixed-period fitting of low-S/N light curves \citep{leleu2}, it may also emerge as a consequence of the respectively varied degrees of atmospheric disruption experienced by near-resonant and nonresonant planets. The density regime occupied by many TTV planets is strongly consistent with the retention of massive gaseous envelopes that were accreted from the protoplanetary disk, while typical RV planets host more tenuous atmospheres that may themselves suggest the loss of a primordial envelope \citep{hadden}, perhaps through disruptive dynamical encounters. Such a paradigm would thus favor a more quiescent evolutionary history for near-resonant planetary systems as compared to their nonresonant counterparts, thereby lying in accord with the notion of distinct formation channels for either configuration as exemplified by the aforementioned turbulent disk and dynamical instability models.

While a suggestive analysis may be performed with extant RV mass measurements, we note that such an inquiry will likely suffer from complex statistical biases generated by the use of heterogenous data and systematic across the sample in question \citep{goyal}, the most heuristically effective comparative analysis would thereby mandate a dedicated, single-instrument RV campaign across both near-resonant and nonresonant configurations. Apart from this potential line of inquiry, we shall note that further insight into such evolutionary histories may be gained from joint consideration of near-resonant and nonresonant configurations in future dynamical modeling efforts.

\section{Acknowledgments} \label{sec:acknowledgments}

We thank Sarah Millholland, Matthias He, Lauren Weiss, Fred Adams, Xi Zhang, Kevin Schlaufman, and Josh Winn for their discussion and insightful commentary on this work, as well as Brandon Radzom, Jessica Ranshaw, Xian-Yu Wang, and Kyle Hixenbaugh for their support during the preparation of this manuscript. We also thank Darin Ragozzine, the anonymous referee, and the statistics editor for their detailed feedback and constructive suggestions, which served to strengthen the quality of this work. S.W. thanks the Heising-Simons Foundation for their generous support.

\bibliographystyle{aasjournal}
\bibliography{main}



\end{document}